\documentclass[aps,nofootinbib,11pt]{revtex4}
\usepackage{amsfonts,amssymb,amsmath}
\def\d{\partial}
\def\x{{\bf x}}

\def\q{{\bf q}}
\def\kB{k_{\rm\scriptscriptstyle B}}

\begin{document}
\title{
  Universal conductivity and central charges
}
\author{Pavel Kovtun$^{a,b}$}
\author{Adam Ritz$^a$}
\affiliation{
  ${}^a$Department of Physics and Astronomy, University of Victoria,
  Victoria, BC, V8P 5C2, Canada\\ 
  ${}^b$Center for Theoretical Physics, Massachusetts Inst. of Technology,
  Cambridge, MA, 02139, USA
}
\date{May 2008}
\begin{abstract}
\noindent
We discuss a class of critical models in $d\geqslant2{+}1$ dimensions
whose electrical conductivity and charge susceptibility are fixed by the
central charge in a universal manner.
We comment on possible bounds on conductivity, as suggested by 
holographic duality.
\end{abstract}
\maketitle


\section{Introduction and summary}
\label{sec:cc}
\noindent
It is not uncommon to find physical systems
which are described by interacting conformal field theories (CFTs).
A simple example is the liquid-gas critical point whose static
correlations are described by the Ising CFT in $d=3$.
Recently, CFTs which are formulated in space-time
(rather than just space) have received attention,
partly due to their appearance in quantum critical
phenomena~\cite{subir-nature,subir-book}.
Such CFTs are relativistic theories, even though their
speed of ``light'' $v$ is not necessarily equal to $3{\times}10^8$m/s.
As a result, charge transport
in these systems at non-zero temperature
obeys simple scaling laws.

At very short distances, the effects of temperature are irrelevant,
and the natural physical questions involve the leading short-distance
singularities of the correlation functions.
On the other hand, at long distances the effects of temperature
become important, and the natural questions are related to
thermodynamics and transport phenomena.
In CFTs, however, short and long distances are related
by a scaling symmetry, and one may anticipate a universal 
relation between the long-distance transport coefficients
and the parameters which describe the short-distance singularities.
Unfortunately, this expectation seems to be quantitatively true
only in $1{+}1$ dimensions. 
The subject of this note is precisely the class of models where such
universal relations between short- and long-distance transport parameters 
extend naturally to any dimension $d>1{+}1$.

At zero temperature,
the (Euclidean) vacuum correlation functions of the energy-momentum tensor
$T_{\mu\nu}$ and a $U(1)$ conserved current $J_\mu$ in a CFT
are fixed to be
\begin{equation}
   \langle J_\mu(x) J_\nu(0) \rangle =
  \frac{k}{x^{2(d-1)}} 
  \frac{1}{\omega_{d-1}^2}
  I_{\mu\nu}\,,
\label{eq:Cmunu}
\end{equation}
\begin{equation}
  \langle T_{\mu\nu}(x) T_{\alpha\beta}(0)\rangle =
  \frac{c}{x^{2d}}
  \frac{1}{\omega_{d-1}^2}
  \left(I_{\mu\alpha} I_{\nu\beta} + I_{\mu\beta} I_{\nu\alpha}
        -\frac2d \delta_{\mu\nu} \delta_{\alpha\beta}\right) \,,
\end{equation}
where $I_{\mu\nu}\equiv\delta_{\mu\nu} - {2x_\mu x_\nu}/{x^2}$,
and $k$ and $c$ are central charges, which are dimensionless constants.
[We use units in which $\hbar{=}v{=}1$ where $v$ is the
speed of ``light'' in the CFT.] 
The factors of $\omega_{d-1}\equiv 2\pi^{d/2}/\Gamma(d/2)$
are inserted for notational convenience.
At non-zero temperature $T$, the equilibrium state is
characterized by pressure $P$, 
as well as by the charge susceptibility
$\chi=\langle Q^2\rangle/(VT)$,
where~$Q$ is the conserved charge associated with the current $J_\mu$,
and we take the thermodynamic limit $V{\to}\infty$.
The susceptibility can be evaluated by introducing
a small chemical potential $\mu$,
so that $\chi(T)=\d \rho/\d\mu|_{\mu=0}$, 
where $\rho(T,\mu)=\langle Q\rangle/V$ is the charge density.
In a CFT, temperature remains the only scale which dictates
\begin{equation}
  P(T)=c'T^d\,,\quad\quad
  \chi(T)=k' T^{d-2}\,,
\end{equation}
where $c'$ and $k'$ are dimensionless constants.
Physically, $c$ and $c'$ provide a measure of
the total number of degrees of freedom in the system,
while $k$ and $k'$ measure the number of charged degrees of freedom.
In two dimensions, $c$ is uniquely related to $c'$ \cite{bcn,affleck}, 
while $k$ is uniquely related to $k'$:
\begin{equation}
  c'=\frac{\pi}{6}c\,,\quad\quad
  k'=\frac{1}{2\pi}k\,,
\label{eq:d=2}
\end{equation}
which means that thermodynamics is uniquely fixed by the central charges.
The reason is that in two dimensional CFTs, the vacuum state is related 
to the thermal state by a symmetry transformation.
We review this argument in the next section.
In $d>2$, the conformal symmetry group is not large enough to
enforce a relation similar to Eq.~(\ref{eq:d=2}), and
therefore thermodynamics is not determined by the central charges.
However, there does exist a large class of CFTs in $d>2$, whose pressure
is determined by the central charge $c$,
resembling the two-dimensional case \cite{kr}.
The crucial property of these models is that they admit a dual description
in terms of classical gravity on a $(d{+}1)$ dimensional
anti-de Sitter space (AdS).
We will show that these CFTs also have the property that their
susceptibility is determined by the central charge $k$,
as for the two-dimensional case.
Namely, we find the following relations:
\begin{equation}
  \frac{c'}{c} = \frac{1}{4\pi^{d/2}}
  \left(\frac{4\pi}{d}\right)^{d} \frac{\Gamma(d/2)^3}{\Gamma(d)}
  \frac{(d{-}1)}{d(d{+}1)}\,,\quad\quad
  \frac{k'}{k} = \frac{1}{2\pi^{d/2}}
  \left(\frac{4\pi}{d}\right)^{d-2} \frac{\Gamma(d/2)^3}{\Gamma(d)}\,.
\label{eq:ckratios}
\end{equation}
Even though the ratios (\ref{eq:ckratios}) are derived
for integer $d\geqslant3$, they can be analytically continued to any
real positive~$d$.
In particular, in $d=2$ they reproduce the universal relations
(\ref{eq:d=2}).

The CFTs which admit a dual gravitational description have many more
universal properties beyond the above relation between
thermodynamics and the central charges.
A surprising feature of these CFTs (and of their relevant
deformations) is that momentum transport in these models is
completely determined by thermodynamics. In particular, their
viscosity is given by $\eta=s/4\pi$ in any dimension \cite{kss,buchel},
where $s=(\d P/\d T)$ is the entropy density.
This is surprising because transport coefficients are typically
determined by the mean-free path even in CFTs \cite{ds},
and are not fixed by thermodynamics.
We will show that charge transport in these models is also
completely determined by thermodynamics.
Namely, we find that the dc electrical conductivity $\sigma$
obeys a similar relation,
\begin{equation}
  \frac{\eta}{s} = \frac{1}{4\pi}\,,\quad\quad
  \frac{\sigma}{\chi} = \frac{1}{4\pi T}\frac{d}{d{-}2}\,.
\label{eq:sigmachi}
\end{equation}
Again, even though the ratio $\sigma/\chi$ was derived for
integer $d\geqslant3$, it can be analytically continued to
any real positive $d>2$.
We will see that the singularity at $d=2$ is precisely what
one expects in $1{+}1$ dimensional CFTs.

The ratio of viscosity to entropy density was conjectured to be
a universal lower bound, saturated by models with a dual gravity
description \cite{kss}.
Motivated by the viscosity bound conjecture,
we discuss similar bounds on conductivity
in relativistic CFTs which are saturated by 
models with gravity duals.

\section{No hydrodynamics in 1+1 dimensions}
\noindent
In this section, we review the argument \cite{hkss}
that $1{+}1$ dimensional
CFTs have no hydrodynamic regime,
and derive the relation (\ref{eq:d=2})
between the susceptibility and the central charge~$k$.

In two dimensions, correlation functions
at zero temperature and finite temperature
can be related to each other \cite{cardy}. 
This is because the transformation
which maps a plane to a cylinder is a conformal transformation,
and therefore is a symmetry transformation in a CFT.
The finite temperature state is obtained by the exponential map
$z=e^{2\pi i\,Tw}$, where $z=x^0{+}ix^1$ represents a point on the plane,
$w=\tau+iy$ represents a point on the cylinder,
and $\tau$ is Euclidean time which is periodic
with period~$1/T$.

For a scalar operator of dimension $\Delta$, a conformal transformation
$\x{\to}\x'$ restricts the two-point correlation function as follows
\begin{equation}
  \langle\phi(\x'_a)\phi(\x'_b)\rangle =
  D(\x_a)^{-\Delta/d}\, D(\x_b)^{-\Delta/d}
  \langle\phi(\x_a)\phi(\x_b)\rangle\,,
\label{eq:phiphi}
\end{equation}
where $D(\x)=|\det(\d\x'/\d\x)|$ is the Jacobian of the
coordinate transformation.
For the above exponential map in $d{=}2$, we have $D(\x)=1/(2\pi T|\x|)^2$,
while the zero-temperature correlator is simply a power-law,
$\langle\phi(\x_a)\phi(\x_b)\rangle=C_\phi/|\x_a-\x_b|^{2\Delta}$.
From the transformation relation (\ref{eq:phiphi}) we find the
finite-temperature correlator of the scalar field,
\begin{equation}
  \langle\phi(\tau,y)\phi(0)\rangle = C_\phi
  \left[
  \frac{(\pi T)^2}
       {\sin[\pi T(\tau{+}iy)]\, \sin[\pi T(\tau{-}iy)]}
  \right]^\Delta\,.
\end{equation}
This expression is periodic under $\tau\to\tau+1/T$
(as it should be), and reduces to the standard power-law result
in the limit $T{\to}0$.
For models with a dual gravitational description,
this form of the correlator was reproduced from small
perturbations of the BTZ black hole in \cite{recipe}.
A similar argument can be applied to the density-density
correlator on the plane in Eq.~(\ref{eq:Cmunu}), which can be written as
\begin{equation}
  C_{00}(x^0,x^1) = -\frac{k}{8\pi^2} 
              \left\{
              \frac{1}{(x^0+ix^1)^2} + \frac{1}{(x^0-ix^1)^2}
              \right\}\,.
\label{eq:C00}
\end{equation}
At finite temperature, we find
\begin{equation}
  C_{\tau\tau}(\tau,y)= -\frac{k}{8\pi^2}
              \left\{ 
              \left[\frac{\pi T}{\sin[\pi T(\tau{+}iy)]}\right]^2 +
              \left[\frac{\pi T}{\sin[\pi T(\tau{-}iy)]}\right]^2
              \right\}\,.
\label{eq:Ctautau}
\end{equation}
Again, this expression is periodic under $\tau\to\tau+1/T$,
and reduces to Eq.~(\ref{eq:C00}) when $T{\to}0$.
The charge susceptibility follows from the thermal
density-density correlator,
\begin{equation}
  \chi = -\frac1T \int\!\!{\rm d}^{d-1}{\bf y}\; 
          C_{\tau\tau}(\tau,{\bf y})\,,
\end{equation}
where the extra minus sign is due to the Euclidean signature.
In $d{=}2$ dimensions, we use the explicit expression
(\ref{eq:Ctautau}), and find $\chi=k/2\pi$, as stated earlier in Eq.~(\ref{eq:d=2}).

The imaginary-time result (\ref{eq:Ctautau}) can be Fourier transformed,
\begin{equation}
  C_{\tau\tau}(\omega_n,q) = \int_0^{1/T} \!\!{\rm d}\tau 
                             \int\!\! {\rm d}y\;
                             C_{\tau\tau}(\tau,y)\;
                             e^{-i\omega_n\tau-iqy}\,,
\end{equation}
where $\omega_n=2\pi n T$ is the Matsubara frequency.
When performing the Euclidean time integration, the domain
can be extended to include the whole real axis, and one picks up
contributions from an infinite sequence of poles in the complex
$\tau$ plane.
For the density-density correlator one finds a simple expression
\begin{equation}
  C_{\tau\tau}(\omega_n,q) = -\frac{k}{2\pi}\;
                             \frac{q^2}{\omega_n^2 + q^2}\,.
\end{equation}
Analytic continuation to real frequency $\omega$
produces the retarded correlator $C_{tt}^{\rm ret}(\omega,q)$
which only has light-cone singularities, but shows no
hydrodynamic modes (as one would find in higher dimensions).
Formal application of the Kubo formula now gives
$$
  \sigma(\omega)={\rm Im} \frac{\omega}{q^2}
                 C_{tt}^{\rm ret}(\omega,q)
                =\frac{k}{2}\, \delta(\omega)\,.
$$
The singularity in the dc limit $\omega{\to}0$ is precisely
what we have in the general result (\ref{eq:sigmachi})
when~$d{=}2$.

\section{Charge susceptibility}
\noindent
We will focus on quantum field theories which admit
a dual description in terms of classical gravity
in Anti-de Sitter (AdS) space within the
AdS/CFT correspondence \cite{magoo}.
For such field theories, a large-volume thermal state
in a $d$-dimensional CFT is described 
by a $(d{+}1)$-dimensional black hole in AdS.
The black hole solution follows from the
Einstein-Maxwell action,
\begin{equation}
  S = \frac{1}{16\pi G} \int\!\! {\rm d}^{d+1}x\, \sqrt{-g}\;
      \left[R+\frac{d(d{-}1)}{L^2}\right] 
     -\frac{1}{4g_{d+1}^2} \int\!\! {\rm d}^{d+1}x\, \sqrt{-g}\;
      F^2\,,
\label{eq:bulk-action}
\end{equation}
where $L^2$ sets the value of the cosmological constant,
and $g_{d+1}^2$ is the $(d{+}1)$-dimensional gauge coupling constant,
which has dimension of (length)$^{d-3}$.
An equilibrium state at finite temperature and density
is described by the Reissner-Nordstrom black hole in AdS.
The thermodynamics of these black holes has been studied extensively,
see for example \cite{cejm1}.
The metric in the thermodynamic limit is given by
\begin{equation}
   ds^2 = \frac{r^2}{L^2} \big(\,-\!\!V(r)dt^2+d\x^2\big) +
             \frac{L^2}{r^2} \frac{dr^2}{V(r)},
\label{eq:RN-AdS-metric}
\end{equation}
where $V(r)= 1-{m}/{r^{d}} + {m_q^2}/{r^{2d-2}}$,
and the boundary is at $r{\to}\infty$.
The parameter $m$ determines the mass of the black hole,
and $m_q$ determines its charge.
The background gauge field is $A_t=\mu-\,C/r^{d-2}$, where
the constant $C$ is related to the charge density of the CFT.
The chemical potential $\mu$ is fixed by the condition
that $A_t$ vanishes at the horizon $r=r_0$, i.e.
\begin{equation}
  \mu = \frac{C}{r_0^{d-2}}\,.
\label{eq:mu}
\end{equation}
The charge density $\rho$ is defined by the variation of the action
with respect to the boundary value of the bulk gauge field
$A_t^{(b)}=A_t(r{\to}\infty)$,
\begin{equation}
  \rho=\frac{\delta S}{\delta A_t^{(b)}}
      = \frac{(d{-}2)C}{g_{d+1}^2 L^{d-1}}\,.
\end{equation}
To find the susceptibility, we need the relation between
$\rho$ and $\mu$ to linear order in $\mu$.
This means that in (\ref{eq:mu}) it suffices to express $r_0$
in terms of temperature when $\mu{\to}0$, and
one finds $T=r_0 d/(4\pi L^2)$.
From the definition of the chemical potential (\ref{eq:mu}) we find
$\rho(T,\mu)=\chi(T)\,\mu$, where the susceptibility%
\footnote{
  The value of the susceptibility can also be deduced from the
  hydrodynamic current-current correlators.
  Namely, one has for the retarded charge density-charge density
  correlation function:
  $  C_{tt}^{\rm ret}(\omega,q) = {\chi D q^2}/(i\omega-D q^2)\,,$ 
  where $D$ is the charge diffusion constant.
  Comparing this with the hydrodynamic correlators 
  found in \cite{pss} for $d{=}4$, and in \cite{herzog} for $d{=}3,6$,
  one finds the susceptibility in $d=3,4,6$
  in agreement with the general expression (\ref{eq:chi}).
}
is
\begin{equation}
 \chi = \frac{(d{-}2)L^{d-3}}{g_{d+1}^2}
        \left(\frac{4\pi}{d}\right)^{d-2} T^{d-2}.
\label{eq:chi}
\end{equation} 
The value of the central charge $k$ can be found from the results of
Freedman et al. \cite{freedman}:
\begin{equation}
  k = \frac{L^{d-3}}{g_{d+1}^2} \frac{\Gamma(d)(d{-}2)}{2\pi^{d/2}\Gamma(d/2)}
      \omega_{d-1}^2\,.
\end{equation}
Comparing with the susceptibility in (\ref{eq:chi}), we find our result for
$k'/k$ in Eq.~(\ref{eq:ckratios}).

\section{Electrical conductivity}
\noindent
The methods of the AdS/CFT correspondence also allow us to
compute the electrical conductivity of CFTs with a dual
gravity description.
Since these models typically do not have dynamical $U(1)$ gauge
fields, we first need to say what we mean by the conductivity.
We imagine gauging a global $U(1)$ symmetry of the theory
with a small coupling $e$, and work to leading order in $e$.
The electrical conductivity is then defined with respect to 
this $U(1)$ gauge field.
To leading order in $e$, the effects of the gauge field
can be ignored, and the electromagnetic response can be determined
from the original theory \cite{CaronHuot:2006te}.
This essentially amounts to sending $J_\mu\to eJ_\mu$, and
a factor of $e^2$ will appear in both the conductivity and the susceptibility.
The conductivity is determined from the real-time current-current
correlation function in thermal equilibrium,
\begin{equation}
  \sigma(\omega)\delta_{ij} = 
  e^2\; {\rm Im}\,\frac{1}{\omega} C_{ij}^{\rm ret}(\omega,\q{=}0)\,.
\label{eq:Kubo}
\end{equation}
Here $C_{ij}^{\rm ret}(\omega,\q)$ is the retarded correlation function of
the global $U(1)$ symmetry currents.
The dc conductivity is $\sigma(\omega{=}0)$.

To evaluate $C_{ij}^{\rm ret}(\omega,\q)$, we use the standard AdS/CFT
recipe of \cite{recipe,pss}, and consider Maxwell fields
propagating on the $(d{+}1)$ dimensional background,
\begin{equation}
  ds^2 = \frac{L^2}{z^2}\left( -f(z)dt^2+d\x^2 + \frac{dz^2}{f(z)} \right)\,,
\end{equation} 
where $f(z) = 1-(z/z_0)^d$,
and the temperature of the CFT is $T= d/(4\pi z_0)$.
This metric is obtained from 
(\ref{eq:RN-AdS-metric}) at $m_q{=}0$,
changing the radial coordinate to $z=L^2/r$.
The bulk action for the Maxwell field is given by (\ref{eq:bulk-action}).
Translation invariance allows us to take
the bulk gauge field proportional to $e^{-i\omega t+ i\q{\cdot}\x}$,
and $\q{=}0$ is sufficient to find the conductivity
using the Kubo formula (\ref{eq:Kubo}).
The component $A_i$ satisfies the equation
\begin{equation}
   u^{d-3}\left[\frac{f(u)}{u^{d-3}}A_i'(u)\right]' +
   \frac{w^2}{f(u)}A_i(u) = 0\,,
\label{eq:Ai-eqn}
\end{equation}
where $u=z/z_0$, and $w=\omega z_0$.
The computation of the retarded correlation function requires the
choice of an outgoing boundary condition at the horizon, 
i.e. $A_i(u) = (1-u)^{-iw/d}\, a(u)$, where $a(u)$ is regular at $u{=}1$.
To find the dc conductivity, we solve the equation for $a(u)$
as a power expansion in frequency, $a(u)=a_0+iwa_0 h(u)+{\cal O}(w^2)$.
For arbitrary dimension $d$, 
the solution for $h(u)$ can be expressed in terms of Gauss'
hypergeometric function, and the integration constants are fixed
by requiring that $h(u)$ vanishes at the horizon. 
The current-current retarded correlation function is evaluated from the
on-shell boundary action,
\begin{equation}
  S = \frac{(L/z_0)^{d-3}}{2 g_{d+1}^2}
      \int\!\frac{\rm d\omega}{2\pi}\frac{{\rm d}^{d-1}q}{(2\pi)^{d-1}}\;
      \frac{ A_i'(\omega,u) A_i(-\omega,u)}{z_0\, u^{d-3}},
\end{equation}
with the implicit limit $u{\to}0$.
The near-boundary expansion for $h(u)$ has the form
$h(u) = h(0)+ \ln(1{-}u)/d + u^{d-2}/(d{-}2)+{\cal O}(u^{2d-2})$
which allows us to read off $C_{ij}^{\rm ret}(\omega,\q{=}0)$ to
leading order in~$\omega$. 
The Kubo formula (\ref{eq:Kubo}) then gives the conductivity,
\begin{equation}
  \sigma = \frac{e^2}{g_{d+1}^2}\left(\frac{L}{z_0}\right)^{d-3}\,.
\end{equation}
On the other hand, the susceptibility (\ref{eq:chi}) can be written as
$ \chi= (e^2/g_{d+1}^2)\, (L/z_0)^{d-3} (d{-}2)/z_0 $
and we arrive at the simple result (\ref{eq:sigmachi}) for the 
conductivity to susceptibility ratio.
For systems in which charge transport proceeds by diffusion,
conductivity is related to the diffusion constant $D$ by the
Einstein relation $\sigma=\chi D$.
Therefore, our result can be interpreted as a remarkably simple
diffusion constant
in~$d$ spacetime dimensions,
\begin{equation}
  D = \frac{1}{4\pi T}\frac{d}{d{-}2}\,.
\label{eq:D}
\end{equation}
One readily verifies that it agrees with the known results
in $d{=}4$ \cite{pss}, and $d{=}3,6$ \cite{herzog}.%
\footnote{
  We were informed by A.~Starinets that he has independently obtained
  Eq.~(\ref{eq:D}) \cite{pc}. }
The electrical conductivity takes a particularly simple form in
$2{+}1$ dimensional CFTs.
In this case the equation (\ref{eq:Ai-eqn}) can easily be solved
for all $\omega$, and one finds a {\em frequency-independent}
optical conductivity \cite{hkss},
$$
  \sigma(\omega)=\frac{e^2}{g_4^2}\,.
$$
It is a peculiar feature of these models that
because of strong quantum fluctuations the optical conductivity
in $2{+}1$ dimensions is frequency-independent, and shows no crossover regime
at $\hbar\omega\sim \kB T$.
It would be very exciting to find two-dimensional materials
which have this property.

\section{A conductivity bound?}
\noindent
We have shown that in all CFTs with a classical gravity dual,
the ratio of electrical conductivity to the static charge
susceptibility is given by a very simple form (\ref{eq:sigmachi}).
Given that in these models the thermodynamics is fixed by the central charges, 
and transport coefficients are fixed by thermodynamics, it follows that
transport coefficients are uniquely fixed by the central charges,
$\eta\sim c\, T^{d-1}$, and $\sigma\sim k\, T^{d-3}$.
Therefore, the ratio of viscosity to conductivity is proportional
to the ratio of the central charges,
\begin{equation}
  \frac{\eta\, e^2}{\sigma T^2} = \frac{8\pi^2(d{-}1)(d{-}2)}{d^3(d{+}1)}\;
                            \frac{c}{k}\;.
\label{eq:eta-sigma}
\end{equation}
If $c$ and $k$ indeed provide a suitable measure of the number
of degrees of freedom in the system,
it is not unreasonable to assume
that the right-hand side of (\ref{eq:eta-sigma}) is bounded
from below%
\footnote{
  There are known counter-examples to the decrease of $c$ and $k$
  along renormalization group
  trajectories in supersymmetric theories \cite{aefj},
  which suggests that $c$ and $k$ do not always unambiguously measure
  the number of degrees of freedom in the theory.
  However, in these examples $k$ corresponds to the $R$-current
  (which is related to the energy-momentum tensor by supersymmetry),
  and therefore $c$ and $k$ are proportional to each other.
 } 
because the number of charged degrees of freedom
must be smaller than the total number of degrees of freedom.
As a result, one could imagine that the conductivity and viscosity obey a
bound of the kind 
$\sigma T^2\leqslant\lambda_d\,\eta\,e^2$,
with some order one constant $\lambda_d$.%
\footnote{
  Alternatively, the ratio of viscosity to conductivity can be expressed
  as the ratio of the $d{+}1$ dimensional gauge coupling constant to 
  the $d{+}1$ dimensional Newton's constant,
  $e^2\eta/(\sigma T^2) = (\pi/d^2)\, (L^2 g_{d+1}^2)/G_N$.
  From the dual gravitational point of view, such an inequality between $c$ and $k$ is
  related to a version of the ``weak gravity conjecture''
  of Ref.~\cite{wgc} in AdS space.
  We thank John McGreevy for pointing out Ref.~\cite{wgc} to us.
}
However, one should keep in mind that
the definition of $\sigma$ (or $\eta$)
involves an arbitrary choice of normalization for
the corresponding current.
[E.g., if the electromagnetic $U(1)$ is chosen as a subgroup
of a larger global symmetry group $G$, this
translates to an arbitrary choice of normalization for the
generators of $G$.]
Therefore, any universal bound on conductivity will more naturally
involve a quantity which is independent of the normalization,
such as $\sigma/\chi$.

The universal relation for $\sigma/\chi$ in (\ref{eq:sigmachi})
looks similar to the universal relation for $\eta/s$: both ratios
become large at weak coupling due to a large mean-free path,
and saturate at strong coupling in CFTs with a dual gravity description.
Alternatively, one may wonder if there could be a 
lower bound for conductivity, similar to the conjectured 
lower bound \cite{kss} for viscosity,
\begin{equation}
  \frac{\sigma}{\chi} \geqslant \frac{\hbar v^2}{4\pi T}\, \frac{d}{d{-}2}\,,
\label{eq:sigma-bound}
\end{equation}
where we have now restored $\hbar$ and the speed of ``light'' $v$.
There is an important difference between the two ratios
in Eq.~(\ref{eq:sigmachi}): 
while $\eta/s=\hbar/4\pi$ only contains $\hbar$
(suppressing the Boltzmann constant),
the corresponding ratio for the conductivity also contains
the speed of ``light''.
The non-relativistic limit corresponds to $v{\to}\infty$,
and therefore the bound (\ref{eq:sigma-bound})
cannot hold in non-relativistic systems.
Within the AdS/CFT correspondence, this expectation is confirmed
by several explicit computations \cite{membrane,mst}
where $\sigma/\chi$
falls below the bound (\ref{eq:sigma-bound}) once conformal
symmetry is broken.
In practice, the speed of ``light'' $v$ is different in different materials,
and is not bounded from below.
Once the speed of light is fixed,
it is not unreasonable to guess that
Eq.~(\ref{eq:sigma-bound}) does represent a lower bound
on conductivity in relativistic CFTs such as the
Wilson-Fisher fixed point in the $O(N)$ model.

\section{Discussion}
\noindent
We have argued that in a large class of CFTs in $d>2$,
there are universal relations between the thermodynamic and 
transport properties, and the central charges which dictate the short distance behaviour of current-current correlators. 
One way of defining
this class of theories is that they possess dual descriptions within AdS at the level of classical gravity and Maxwell electrodynamics.
For example, this universality determines the shear viscosity
$\eta$ and ``electrical" conductivity $\sigma$ in terms of the corresponding
central charges and naturally leads to a conjectured bound on conductivity 
in physical systems, given in Eq.~(\ref{eq:sigma-bound}),
in analogy with the well-studied viscosity bound conjecture.

It is natural to ask about the regime of validity,
or alternatively the constraints on the CFTs which may enter 
such universality classes. 
Indeed, the analysis we have performed using the AdS/CFT duality 
required the validity of a classical gravity approximation, and thus
some kind of a large-$N$ limit.
On the gravity side of the duality, universality follows 
from the uniqueness of the lowest dimension operator which determines
the dynamics of the metric and/or the gauge field dual to the current in question,
i.e. the uniqueness of the Einstein-Hilbert  and Maxwell actions respectively. 
From this point of view, once we move to finite $N$, 
it appears that a large number of higher
derivative corrections will also be required, 
thus limiting the possibility for universal behaviour.
Nonetheless, it would be interesting
if additional symmetries on the bulk side could constrain 
the possible classes that might arise. 
Some hint in this direction is provided by the black hole solutions 
in string theory beyond the leading classical Einstein term \cite{sen}.

With these issues in mind, it is clearly useful to have a concrete example
with which to contrast the general holographic results.
We will consider the 3-dimensional $O(N)$ model at large $N$
with fields $\phi^\alpha$, $\alpha=1,..,N$ 
subject to a constraint $\phi^\alpha\phi^\alpha=1$. 
In this system, the ratio of $c'/c$ was computed in the large-$N$ limit by Sachdev \cite{sachdev}, with the result,
\begin{equation}
 \left(\frac{c'}{c}\right)_{\!\!O(N)} = \frac{8\zeta(3)}{15\pi} \approx 0.2041,
\end{equation}
which differs by only a few percent from the holographic answer, 
$c'/c =\pi^3/162 \approx 0.1914$ \cite{kr}.
A similar comparison is possible for the ratio $k'/k$,
where a natural variant of the vector
current studied above lies in the adjoint, 
$J_\mu^{\alpha\beta} = 
 (\phi^\alpha\d_\mu\phi^\beta-\phi^\beta\d_\mu\phi^\alpha)$,
and we can write
\begin{equation}
 \langle J^{\alpha\beta}_\mu(x) J^{\gamma\delta}_\nu(0) \rangle = 
  \frac{k}{x^{4}} \left( \delta_{\mu\nu} - \frac{2x_\mu x_\nu}{x^2}\right)
  \frac{
  \delta^{\alpha\gamma}\delta^{\beta\delta} - 
  \delta^{\alpha\delta}\delta^{\beta\gamma}
  }
  {(4\pi)^2}.
\end{equation}
A straightforward calculation of the central charge at large $N$
leads to  $k=2$ \cite{petkou},%
\footnote{
   Alternatively, one can think of the central charge $k$ as the
   dynamical conductivity in the regime $\omega \gg T$.
   For high frequencies, we can use the zero-temperature
   correlation functions (\ref{eq:Cmunu}), and obtain
   $\sigma(\omega{\to}\infty) = e^2 k/32$.
  }
while the charge susceptibility in this case was computed 
by Chubukov et al. \cite{csy}. 
Using these results, we find at large~$N$,  
\begin{equation}
   \left(\frac{k'}{k}\right)_{\!\!O(N)} = 
   \frac{\sqrt{5}}{2\pi} \ln\left(\frac{\sqrt{5}+1}{2}\right)  \approx 0.1713,
\end{equation}
which differs by about 24\% from the result $k'/k = \pi/24 \approx 0.1309$
for models with gravitational duals in AdS.
On the other hand, the conductivity in the $O(N)$ model is large,
$\sigma/\chi$ is ${\cal O}(N)$ \cite{subir-book},
reflecting the fact that the model becomes weakly coupled at large $N$.
Therefore, the comparison with the $O(N)$ model in $d=3$ provides an
example of a situation where two systems
have very similar static thermodynamic properties, but
vastly different transport properties.

As another aspect of the constraints defining these universality classes,
it is possible to consider even more restrictive models.
Namely,  for theories with four supercharges, e.g.
${\cal N}{=}1$ supersymmetry in $d=4$, there 
is a global U(1) $R$-symmetry whose current lies in the same 
supermultiplet as the energy momentum tensor. 
If we use the $R$-current to determine $k$, it follows that $c$ and $k$
are not independent, and specifically that the ratio $c/k$ 
depends only on the dimension~$d$. 
Therefore, for these systems the viscosity 
and the $R$-current conductivity are related by a simple dimension-dependent constant
as given in Eq.~(\ref{eq:eta-sigma}).
As an interesting corollary,
in such models the thermodynamic properties, 
the viscosity, and the conductivity
are fully determined by a single number, the central charge~$c$.

In this paper we focused on a universality class of models
in $d>2$ spacetime dimensions whose thermodynamic properties, 
transport coefficients, and central charges 
were all related to each other. 
We would like to conclude by pointing out that there is more than one
such universality class, and that universal relations such as
Eq.~(\ref{eq:ckratios}) are not restricted to models which admit 
a dual gravitational description.
Rather, they can arise as a natural consequence of a large-$N$ limit.
As examples, we have in mind pairs of ``parent'' and ``daughter'' theories where the
daughter is obtained by projection onto a sector invariant under a global discrete 
symmetry, such as those studied in \cite{asv,kuy}. As a simple example, consider a parent $U(kN)$ gauge theory with matter fields
in the adjoint representation, and project out by a global ${\bf Z}_k$ symmetry
to form a daughter theory with $U(N)^k$ gauge group, and
matter fields in the bi-fundamental representation.
It was shown in \cite{kuy,kuy2} that the parent and daughter theories
are completely equivalent in the ${\bf Z}_k$-invariant sector at large $N$, provided
that the ${\bf Z}_k$ symmetry is not spontaneously broken.
In this and similar examples, since the energy-momentum tensor does not carry any global charges
it follows that the correlation functions of $T_{\mu\nu}$
in the parent and daughter theories are proportional to each other.
For example, 
viscosity is proportional to the two-point correlation function of
$T_{\mu\nu}$, and thus 
all daughters have the same $\eta/s$ ratio.
For CFTs, all daughters would also have the same $c'/c$ ratio.
One can say that the universality class consists of
all daughter theories, which 
(even though they may have different global symmetries and
contain matter fields in different representations)
share the same universal ratios, similar to 
Eqs.~(\ref{eq:ckratios}) and (\ref{eq:sigmachi}).
In this simple example, the existence of such a universality class
requires a large-$N$ limit in the field theory, but
assumes nothing about strong coupling, supersymmetry, 
conformal symmetry, or a dual gravity description.
The implications of such large-$N$ equivalences for universal transport
properties deserve investigation, and we plan to return to them in the future.

\acknowledgments
\noindent
We thank Hong Liu, John McGreevy, and Subir Sachdev for helpful conversations,
and Sean Hartnoll for comments on the manuscript.
This work was supported in part by NSERC of Canada and
by funds provided by the U.S. Department of Energy
(D.O.E.) under cooperative research agreement DE-FG0205ER41360.

\end{document}